\documentstyle[twoside,fleqn,espcrc2,psfig]{article}

\newcommand{\AmS}{{\protect\the\textfont2
  A\kern-.1667em\lower.5ex\hbox{M}\kern-.125emS}}

\title{Nonperturbatively Improved Hadron Spectroscopy Near the 
       Continuum Limit}

\author{
        T. Mendes\address{Gruppo APE,
        Dipartimento di Fisica,
        Universit\`a di Roma ``Tor Vergata'',
        Via della Ricerca Scientifi- ca 1,
        00133 Roma, ITALY}
}
       
\begin{document}

\begin{abstract}
We report the results of our quenched lattice simulations 
of the Wilson action with a nonperturbatively determined
clover term at $\beta=6.2$ and compare them with those 
of the standard Wilson action at the same $\beta$ value.
\end{abstract}

\maketitle

Simulations in full QCD are extremely slow, and are still carried
out far from the continuum limit.
A possible solution for this problem is the use of improved actions,
for which the simulation gives the continuum limit at higher values of
the cutoff $a$, reducing computational costs. An important step
in the improvement program was taken recently by the Alpha
collaboration, with the
{\em nonperturbative} determination of the coefficient $c_{SW}$
in the {\em clover} action \cite{Luscher,Luscher2}.

We present here the results of our study of hadron spectroscopy
using the clover-improved Wilson action
(see \cite{Luscher} for notation)
\begin{eqnarray}
S &=& S_G \,+\, S_W \nonumber \\
             & & \quad \,+\,  c_{SW}\,a^5\,\frac{i}{4}
\sum_x {\overline \Psi}(x)\,\sigma_{\mu\nu}{\hat F}_{\mu\nu}\Psi(x)
\;\mbox{.}
\nonumber
\end{eqnarray}
The nonperturbative expression of $c_{SW}$ is given in terms of 
the coupling $\beta$ in \cite{Luscher2}.
We consider lattice volume $24^3\times 48$ and coupling $\beta = 6.2$.
We thus take $c_{SW}= 1.61375065 $. 
We also consider the unimproved case ($c_{SW}=0$) for comparison.
For the improved case we consider the following values for the hopping 
parameter $\kappa$:
0.1240, 0.1275, 0.1310, 0.1340, 0.1345, 0.1350, 0.1352.
For the $c_{SW}=0$ case we consider:
0.1350, 0.1400, 0.1450, 0.1506, 0.1510, 0.1517, 0.1526.
The simulations were carried out on the APE100 computer at {\em Tor Vergata}.

Our statistics come from 50 quenched gauge configurations,
generated by a hybrid overrelaxation algorithm
with each update corresponding to a heat-bath sweep followed by 3 
overrelaxation sweeps. 
The configurations are separated by 1000 updates.

The inversion of the fermion matrix is performed using the 
stabilized biconjugate gradient algorithm \cite{BiCGStab}.
We restart the inversion from the current solution every 100 iterations,
in order to reduce accumulation of roundoff errors \cite{refresh}.
We employ point-like sources.

We sum fermion propagators over the space directions $x,y,z$ for
sites within blocks of side 3, and then store the result. We then
form hadron correlations from these ``packed'' propagators. This procedure
becomes exact in the limit of an infinite number of configurations, 
since it corresponds to including gauge-noninvariant terms 
in the computation of hadron correlations (in our case we have checked that 
the errors thus introduced are negligible).
This corresponds to gaining a factor $3^3$ in storage, 
and has enabled us to have all quark propagators stored simultaneously. 
In this way we can form hadron correlations from nondegenerate combinations
of quark flavors, as done in \cite{Sharpe}.

\vskip 3mm
Within the improvement program one can determine the critical value
$\kappa_c$ using the unrenormalized current quark mass or
``Ward identity'' mass, defined 
through the improved PCAC relation as \cite{Luscher}
$$ m_{WI} \; \equiv \;
\frac{< \partial_{\mu} \{ {A_{\mu}^{(bare)}} \,+\,
c_A \, a \, \partial_{\mu}{P^{(bare)}}\}{\cal O}>}{2<{P^{(bare)}}{\cal O}>}
\;\mbox{.}
$$
Linear extrapolation to the limit of $\,m_{WI}=0\,$ provides a much more 
stable fit for the determination of $\kappa_c$ than the conventional 
fit of pseudo-scalar meson masses to the limit of zero pion mass.
Our results for $\kappa_c$ are given below.
\begin{eqnarray}
{\bf \mbox{improved case:}} \qquad & & \nonumber \\
\mbox{from} \quad m_{WI} = 0 \; & & \kappa_c = 0.135802(6) \nonumber \\
\mbox{from} \quad M_{PS}^2 = 0 \; & & \kappa_c = 0.135861(19) \nonumber \\
{\bf c_{SW}=0 \;\;\mbox{case:}} \qquad \;\; & &  \nonumber \\
(\mbox{from} \quad M_{PS}^2=0) & & \kappa_c = 0.153307(19) \nonumber 
\end{eqnarray}

Once the value for $\kappa_c$ has been determined, one can obtain chiral
extrapolations from plots of hadron masses as linear functions of the bare
quark mass, defined as 
$ m_q(\kappa) \,\equiv\, (1/\kappa\,-\,1/{\kappa_c})/2$,
or as linear functions of the measured squared pseudoscalar 
mass $M_{PS}^2(\kappa)$.
In the improved case one can define the modified or
``improved'' bare quark mass \cite{Luscher} by
$ {\widetilde m_q}(\kappa) \;\equiv\; m_q(\kappa)\,[1\,+\,b_m\,m_q(\kappa)] $
(Note that ${\widetilde m_q}$ is the renormalized mass with $\,Z_m=1\,$)
The improvement coefficient $b_m$ has been determined nonperturbatively 
\cite{Giulia}:
$ b_m \;=\; -0.60\;\mbox{.} $
Note that since we have nondegenerate flavor combinations we use averages
of the masses defined above, e.g. for a meson corresponding
to the flavors $\kappa_1$ and $\kappa_2$ we define
$  m_q(\kappa_1,\kappa_2) \;\equiv\; 
[m_q(\kappa_1)\,+\, m_q(\kappa_2)]/2$.

\vskip -0.8cm
\protect\footnotesize
\begin{table}[htb]
\addtolength{\tabcolsep}{2mm}
\begin{center}
\begin{tabular*}{7.5cm}{ccc}
\hline
$a^{-1}$ from & IMPROVED  & $c_{SW}=0$ \\
\hline
$ M_{\rho} $   & 2475(231)   & 2759(223)    \\
\hline
$ M_{K^{*}}^{(\rm{I})}$ & 2526(122)  & 2839(115)  \\
\hline
$ M_{\Phi}^{(\rm{I})}$ & 2575(42)  & 2915(39)  \\
\hline
$ M_{\Phi}^{(\rm{II})}$ & 2624(54)  & 2945(44)  \\
\hline
\end{tabular*}
\parbox{7.2cm}{
\vspace{0.2cm}
\caption{\label{tab:am1}
\vskip -0.8cm
\hskip 1.25cm
\protect\small
: Values of $a^{-1}$ in MeV. }
}
\vspace{-1cm}
\end{center}
\end{table}
\normalsize

\protect\footnotesize
\begin{table}[htb]
\addtolength{\tabcolsep}{1mm}
\begin{center}
\vspace{-1.3cm}
\begin{tabular*}{7.5cm}{cccc}
\hline
                & Exp. & IMPROVED  & $c_{SW}=0$ \\
\hline
$ J_{K^{*}}$ & 0.487 & 0.40(14)  & 0.40(9) \\
\hline
$ J_{\Phi} $ & 0.557 & 0.44(16)  & 0.45(10) \\
\hline
\end{tabular*}
\parbox{7.2cm}{
\vspace{0.2cm}
\caption{\label{tab:J}
\vskip -0.8cm
\hskip 1.25cm
\protect\small
: Values of the quantity $J$ and comparison with experiment. }
}
\vspace{-1cm}
\end{center}
\end{table}
\normalsize

In Table \ref{tab:am1} we give our values of the inverse lattice
spacing $a^{-1}$ coming from ratios of the vector meson mass
$M_V$ over its experimental value at different flavor combinations.
We determine the strange-quark mass $m_s$ in lattice units
from two methods: a fit of $(M_{PS}/M_{V}^{\chi})^2$ (method \rm{I})
and a fit of $(M_{PS}/M_{V})^2$ (method \rm{II}) as functions
of the quark mass, interpolating to the experimental values
of these ratios. 
(Here $M_{V}^{\chi}$ denotes the chiral extrapolation
of $M_V$.) For the improved case we use ${\widetilde m_q}$ 
as the bare quark mass, while in the unimproved case we use $m_q$.
In Table \ref{tab:J} are the values of the quantity
$J\equiv M_V\,(dM_V/dM_{PS}^2)$ obtained
from $(M_K/M_{K^{*}})^2$ vs. $M_{PS}^2 $.

\begin{figure}
\vspace{-0.5cm}
\centerline{\psfig{figure=oct_mpio2.axis.ps,height=3in}}
\vspace{-2cm}
\end{figure}
\begin{figure}
\centerline{\psfig{figure=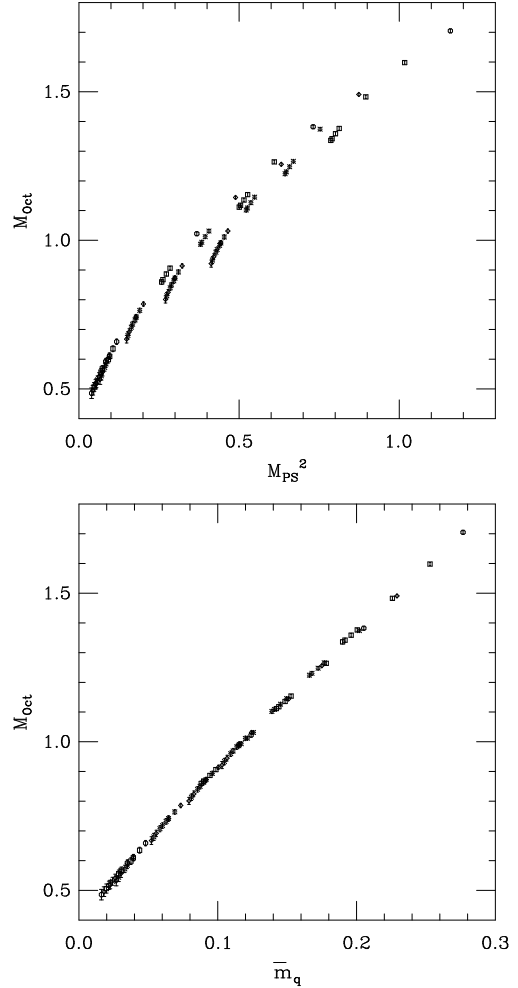,height=3in}}
\vspace{-1.5cm}
\caption{~Octet mass in terms of (i)$M_{PS}^2$ (ii)${\widetilde m_q}$.}
\label{fig:oct}
\vspace{-0.8cm}
\end{figure}

 
\vskip 3mm
All our baryon mass values for the improved case come from linear fits
of the masses as functions
of the improved quark mass ${\widetilde m_q}$,
except for the $\Sigma - \Lambda$ mass splitting, where
we used a quadratic fit.
(Similarly for the $c_{SW} = 0$ case, using as variable
$m_q$.)
 
Note that we have defined the quantity
$ M_{Oct} \equiv (M_N + M_{\Lambda})/2 $
(this combination was chosen so that the resulting mass is
flavor-symmetric). For this quantity the advantage of using the
improved quark mass is clearest, see Fig. \ref{fig:oct}.
(A similar improvement is observed also with respect to plots
using the unimproved bare quark mass $m_q$.)
 
\begin{figure}
\vspace{-0.5cm}
\centerline{\psfig{figure=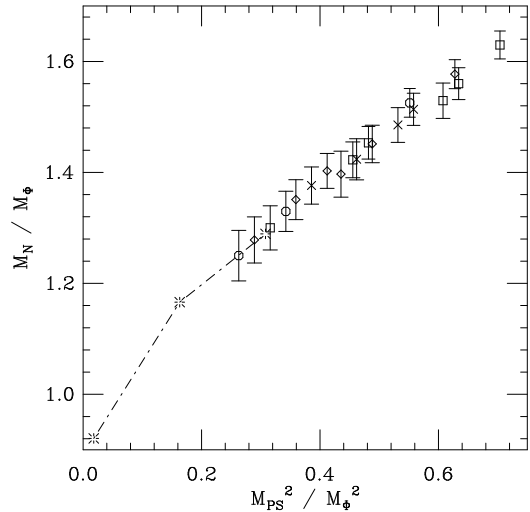,height=3.2in}}
\vspace{-1.4cm}
\caption{~APE plot for the nucleon mass.}
\label{fig:APE}
\vspace{-0.5cm}
\end{figure}

We show an APE plot in Fig. \ref{fig:APE} for the improved case.
Note that we divide at each point by $M_V$
interpolated to the strange-quark mass, corresponding to $M_{\Phi}$
(i.e. a constant value). Experimental
points in these figures correspond to $M_N$, $M_{\Sigma}$ and $M_{\Xi}$
and appropriate meson masses.
In Table \ref{tab:masses1} we give our mass values in MeV.
Table \ref{tab:masses2} contains data in lattice units for the improved case
(diagonal flavor combinations only).

\protect\footnotesize
\begin{table}[htb]
\addtolength{\tabcolsep}{0mm}
\begin{center}
\vspace{-0.5cm}
\begin{tabular*}{7.5cm}{cccc}
\hline
Particle & Exp. value & $c_{SW} = 0 $ & IMPROVED \\
\hline
$ M_{N} $              & $ 939      $ & $ 977(115) $  & $ 953(112) $ \\
\hline
$ M_{\Lambda} $        & $ 1115.7   $ & $ 1142(113) $ & $ 1127(115) $ \\
\hline
$ M_{\Delta} $         & $ 1232     $ & $ 1382(172) $ & $ 1307(160) $ \\
\hline
$ M_{Oct} $            & $ 1152.55  $ & $ 1138(113) $ & $ 1141(113) $ \\
\hline
$ M_{\Delta-N} $       & $ 293      $ & $ 381(90) $ & $ 366(86) $ \\
\hline
$ M_{\Sigma^{*}-Oct} $ & $ 232.45   $ & $ 320(70) $ & $ 295(66) $ \\
\hline
$ M_{\Sigma-\Lambda} $ & $ 73.7     $ & $ 50(39)  $ & $ 59(25) $ \\
\hline
\end{tabular*}
\parbox{7.5cm}{
\vspace{0.2cm}
\caption{\label{tab:masses1}
\vskip -0.8cm
\hskip 1.25cm
\protect\small
: Baryon masses in MeV and comparison with experiment.}
}
\vspace{-1cm}
\end{center}
\end{table}
\normalsize

\vskip -0.5cm
\protect\footnotesize
\begin{table}[htb]
\addtolength{\tabcolsep}{0mm}
\begin{center}
\vspace{-0.8cm}
\begin{tabular*}{7.5cm}{cccc}
\hline
$\kappa $ & ${M_{PS}}^2$ & $M_V$  & $M_N $\\
\hline
0.1240 & 1.15951 & 1.1044(18) & 1.7046(54) \\ \hline
0.1275 & 0.73101 & 0.8922(19) & 1.3821(50) \\ \hline
0.1310 & 0.367794 & 0.6595(22) & 1.0220(55) \\ \hline
0.1340 & 0.117867 & 0.4379(42) & 0.6588(91) \\ \hline
0.1345 & 0.0833452 & 0.3995(59) & 0.5929(100) \\ \hline
0.1350 & 0.0517096 & 0.3660(133) & 0.5169(141) \\ \hline
0.1352 & 0.0397285 & 0.3543(225) & 0.4859(177) \\ \hline
\end{tabular*}
\parbox{7.5cm}{
\vspace{0.2cm}
\caption{\label{tab:masses2}
\vskip -0.8cm
\hskip 1.25cm
\protect\small
: Masses in lattice units. }
}
\vspace{-1cm}
\end{center}
\end{table}
\normalsize

With our present statistics the main effect of the
nonperturbative improvement observed on hadron masses is
a smaller spread in the value of the lattice spacing extracted
from mesons with and without strange quarks.
We have also shown how the use of the improved quark mass 
turns the rough behavior of the dependence of the octet baryon 
mass upon quark masses into a smooth one.

We are currently increasing our statistics \cite{atti}, and we plan to extend
our simulations to the case with bermions \cite{bermion}, an approximate
method that allows for an estimate of dynamic quark effects.

\end{document}